\documentclass [a4paper,12pt]{article}
\pdfoutput=1
\usepackage[english]{babel}
\usepackage{graphicx,amsmath,slashed,caption}

\newcommand{\be}{\begin{equation}}
\newcommand{\ee}{\end{equation}}

\newcommand{\HH}{{\cal H}}
\newcommand{\n}{\hat n}
\newcommand {\apgt} {\ {\raise-.5ex\hbox{$\buildrel>\over\sim$}}\ }
\newcommand {\aplt} {\ {\raise-.5ex\hbox{$\buildrel<\over\sim$}}\ }

\begin{document}
\def\thefootnote{\fnsymbol{footnote}}

\begin{center}
\Large{\textbf{The CMB bispectrum in the squeezed limit}} \\[0.5cm]
 
\large{Paolo Creminelli$^{\rm a}$, Cyril Pitrou$^{\rm b}$ and Filippo Vernizzi$^{\rm c}$}
\\[0.5cm]

\small{\textit{$^{\rm a}$ Abdus Salam International Centre for Theoretical Physics\\ Strada Costiera 11, 34151, Trieste, Italy}}\\[0.2cm]
\small{\textit{$^{\rm b}$ Institute of Cosmology \& Gravitation, University of Portsmouth\\ Portsmouth PO1 3FX, United Kingdom}}\\[0.2cm]
\small{\textit{$^{\rm c}$ CEA, IPhT, 91191 Gif-sur-Yvette c\'edex, France \\  CNRS, URA 2306, F-91191 Gif-sur-Yvette, France}}

\end{center}

\vspace{.8cm}

\hrule \vspace{0.3cm}
\noindent \small{\textbf{Abstract}\\
The CMB bispectrum generated by second-order effects at recombination can be calculated analytically when one of the three modes has a wavelength much longer than the other two and is outside the horizon at recombination. This was pointed out in \cite{Creminelli:2004pv} and here we correct their results.  We derive a simple formula for the bispectrum,  $f_{\rm NL}^{\rm loc} = -  (1/6+ \cos 2 \theta) \cdot (1- 1/2 \cdot d \ln (l_S^2 C_{S})/d \ln l_S )$, where $C_S$ is the short scale spectrum and $\theta$ the relative orientation between the long and the short modes. This formula is exact and takes into account all effects at recombination, including recombination-lensing, but neglects all late-time effects such as ISW-lensing. The induced bispectrum in the squeezed limit is small and will negligibly contaminate the Planck search for a local primordial signal: this will be biased only by $f_{\rm NL}^{\rm loc}\approx-0.4$.  The above analytic formula includes the primordial non-Gaussianity of any single-field model. It also represents a consistency check for second-order Boltzmann codes: we find substantial agreement with the current version of the CMBquick code. 
\\
\noindent
\hrule
\def\thefootnote{\arabic{footnote}}
\setcounter{footnote}{0}


\vspace{0.5cm}
\def\thefootnote{\arabic{footnote}}
\setcounter{footnote}{0}

\section{Introduction and main results}

The Planck satellite \cite{planck} is currently taking data and has as one of its main objectives the exploration of non-Gaussianities in the $f_{\rm NL} \sim $ few range. The primary interest is of course in primordial non-Gaussianities generated in the early Universe, but for such small values of  $f_{\rm NL}$ all second-order effects connecting the initial conditions to the observed CMB are potentially relevant. The theoretical control of all these effects is still not completely satisfactory. Though the complete second-order equations have been studied, see for example \cite{Bartolo:2004ty,Boubekeur:2009uk,Bartolo:2005kv,Bartolo:2006cu,Bartolo:2006fj,Pitrou:2008ak,Pitrou:2008hy,Bartolo:2008sg,Khatri:2008kb, Khatri:2009ja,Senatore:2008vi,Senatore:2008wk,Nitta:2009jp,Pitrou:2010sn,Beneke:2010eg}, getting the full bispectrum induced by second-order effects requires to resort to a numerical code. To our knowledge, the only publicly available code including all second-order effects is CMBquick \cite{cyril}. This is hardly satisfactory as the complexity of the bispectrum calculation and the importance of the result would require a cross-check among different programs. 

However, as pointed out in \cite{Creminelli:2004pv} the full bispectrum calculation becomes quite easy in the limit in which one of the modes is out of the Hubble radius at recombination, while the other two have a much shorter wavelength. The reason for the simplification is intuitive: a mode which is out of the Hubble radius cannot affect any physical process, so that the effect of the long mode will only intervene in the way the short-scale 2-point function is eventually observed. The purpose of this paper is to review and correct the results of \cite{Creminelli:2004pv} and to compare the obtained bispectrum with the full calculation of the second-order Boltzmann code CMBquick \cite{cyril}, the publicly available Boltzmann code created and maintained by  one of the authors of this paper.  

To calculate the bispectrum we have to evaluate the temperature anisotropy at second order (the explicit calculation in the fluid approximation is reviewed in appendix~\ref{app:CMB}), in the presence of a short and a long mode, much longer than the Hubble radius at recombination. The long mode has no physical effect at recombination: it just acts as an unobservable coordinate redefinition, as we will explain in Section \ref{sec:coord}, while recombination takes place perturbed---at linear order---by the short mode. If the long mode were even longer than the present Hubble radius this would be the end of the story, as its presence would not leave any effect whatsoever. But for the calculation of the bispectrum, we are interested in the situation in which the mode eventually comes back into the Hubble radius and we have to compare  different large portions of the sky with the same physics at recombination, but modulated by a long mode. 

The long mode gives rise to various effects that will be discussed in Section \ref{sec:long}. First of all, it gives a long-wavelength modulation of the temperature so that, even if the short-scale perturbations have the same dynamics in the various regions, the perturbations are defined with respect to a different average temperature in different directions of the sky. This gives the same effect as a local primordial non-Gaussianity, with $f_{\rm NL}^{\rm loc} = -1/6$, and it was missed in \cite{Creminelli:2004pv}.\footnote{The presence of this term was shown in the limit where all the modes are out of the Hubble radius in \cite{Bartolo:2004ty} and was later understood as a consequence of the local redefinition of the average temperature in \cite{Boubekeur:2009uk}.}  The long mode also gives rise to geometrical effects: the relation between a physical scale at recombination and a coordinate length will be modulated by the long mode and this will induce an angular stretching of the 2-point function in different regions of the sky. Thus, the effect of the long mode will be proportional to the derivative of the 2-point function with respect to the angular separation in the sky. Finally, lensing close to the last scattering surface will also displace the observed 2-point function and part of this effect will also depend on the relative orientation between the long and the short modes. 

We will get the final bispectrum in Section \ref{sec:bispectrum}. We can write it as a function of the power spectra of the long and short modes, $C_{l_L}$ and $C_{l_S}$,  and the relative angle between them, $\theta$, in the form
\be
\boxed{B_{l_L l_S l_S}=  C_{l_L} C_{l_S} (1+  6 \cos 2 \theta)\left( 2 - \frac{d \ln (l_S^2 C_{l_S})}{d \ln l_S}  \right) }  \;. \label{formula1}
\ee
This formula is exact in the squeezed limit as it includes all second-order effects at last scattering (including the early ISW effect and the effect of anisotropic stresses). Additional contributions to the bispectrum will come at much lower redshifts and are therefore physically distinct and can, in principle, be observationally separated. When the Universe starts accelerating, a large ISW-lensing signal is induced \cite{Seljak:1998nu,Goldberg:1999xm,Hanson:2009kg,Lewis:2011fk}, while when structures begin to form a host of non-linear effects takes place, especially on short scales, which are far from Gaussian (see for example \cite{Cooray:1999kg,Mangilli:2009dr}). Notice that the formula above takes into account the correlation of lensing (induced by the long mode) with the temperature fluctuation at last scattering: this is important as there is a relevant cancellation \cite{Creminelli:2004pv} between the isotropic part of lensing and the redefinition of spatial scales discussed above. 

The comparison of our analytical result with the current version of CMBquick \cite{cyril} is successful (Section \ref{sec:comparison}), with some residual discrepancy most likely coming from the numerics, especially in the treatment of neutrinos. Notice that for the code the check is far from trivial: though in the squeezed limit the understanding of the CMB bispectrum is much easier, in the code the result is not dominated by few simple effects. On the contrary, all second-order effects must conspire to reproduce the simple formula above.

We will show that eq.~\eqref{formula1} already takes into account all the (usually small) primordial non-Gaussianity of single-field models. From this point of view, it can be seen as a clear observational prediction of all these scenarios and as an observational extension of the Maldacena consistency relation \cite{Maldacena:2002vr,Creminelli:2004yq,Cheung:2007sv}. 

The bispectrum that we obtain is rather small and unlikely to be a worrisome source of contamination for the forthcoming Planck analysis. This is quantified in Section \ref{sec:contamination}, where we find that second-order effects at recombination would appear as a small bias to a search of a local primordial non-Gaussianity by $f_{\rm NL}^{\rm loc}\approx-0.4$. This implies that, in the squeezed region, where the primordial local signal is peaked, we have an analytical control of the background induced by second-order effects. Conclusions are drawn in Section \ref{sec:conclusions}.


\section{A long mode as a coordinate transformation}
\label{sec:coord}

In this section we will discuss how second-order perturbations, in the limit in which one of the modes is out of the Hubble radius, can be obtained simply by a coordinate redefinition of the first-order results.
We will start showing that it is possible to rewrite locally a perturbed FRW metric as an unperturbed one by reabsorbing a very long wavelength mode---a mode outside the Hubble radius---with a coordinate transformation. 

We start from a linearly perturbed metric  in Newtonian gauge and we consider only scalar perturbations. The metric reads
\be
ds^2 = a^2 (\eta) \left[ - (1+ 2 \Phi_L) d \eta^2 + (1-2 \Psi_L) \delta_{ij} dx^i dx^j \right] \;, \label{metric_1}
\ee
where $\Phi_L$ and $\Psi_L$ are very long wavelength modes, thus practically independent of the position. 
We consider the following coordinate transformation \cite{Weinberg:2003sw}
\be
\begin{split}
\tilde \eta &= \eta + \epsilon(\eta) \;,  \\
\tilde x^i &= x^i ( 1 - \lambda )\;, \label{ct}
\end{split}
\ee
where $\epsilon$ is an arbitrary function of time and $\lambda$ is an arbitrary constant. This is the most generic coordinate transformation of an unperturbed FRW solution which leaves the metric in Newtonian gauge.  In the new coordinates the metric takes the form \eqref{metric_1} with potentials given by
\begin{align}
\tilde \Phi_L = \Phi_L - \epsilon' - \HH \epsilon \;, \\
\tilde \Psi_L = \Psi_L - \lambda + \HH \epsilon\;.
\end{align}
Thus, we can reabsorb the long-wavelength mode by  choosing $\epsilon$ and $\lambda$ such that $\tilde \Phi_L =0$ and $\tilde \Psi_L=0$, i.e.
\be
\Phi_L =  \epsilon'  +\HH \epsilon \;, \qquad \Psi_L =  \lambda - \HH \epsilon\;. \label{conds}
\ee

As explained in \cite{Weinberg:2003sw}, not all the choices of $\epsilon$ and $\lambda$ give rise to the long-wavelength limit of a physical mode in Newtonian gauge.
Indeed, $\epsilon$ and $\lambda$ must satisfy certain conditions imposed by those Einstein's equations that vanish in the limit $k_L\to 0$, where $ k_L$ is the wavevector associated to $\Phi_L$ and $\Psi_L$. These conditions are 
\be
(\HH' - \HH^2) v = (\Psi' + \HH \Phi)\;, \label{cond_ct_1}
\ee
where $v$ is the velocity potential, and 
\be
\Phi = \Psi - 8 \pi G \delta \sigma\;, \label{cond_ct_2}
\ee
where $\delta \sigma$ is the anisotropic stress, defined by writing the spatial part of the stress-energy tensor as $T_{ij} = g_{ij} p + \partial_i \partial_j \delta \sigma$. Using eqs.~\eqref{conds} and the first condition \eqref{cond_ct_1} one finds that $v = -\epsilon$ and that $\lambda$ simply coincides (up to a sign) with the comoving curvature perturbation $\zeta \equiv -\Psi + \HH v$ (\footnote{In \cite{Weinberg:2003sw} and other references this variable is more commonly called $\cal R$. Here we use the notation of \cite{Maldacena:2002vr}.}) in the limit $k_L \to 0$,
\be
\label{eq:lambdazeta}
\lambda = - \zeta_L\;.
\ee
Using the above equation the second condition, eq.~\eqref{cond_ct_2}, reads 
\be
\epsilon' + 2 \HH \epsilon = - \zeta_L - 8 \pi G \delta \sigma_L\;,
\ee
which can always be satisfied by a suitable choice of $\epsilon$. In particular, integrating this equation yields
\be
\epsilon = - \frac{1}{a^2} \int a^2 (\zeta_L + 8 \pi G \delta \sigma_L) d \eta\;. \label{eps}
\ee

For instance, integrating the above equation in the absence of anisotropic stress, one finds for radiation dominance ($a \propto \eta$) $\epsilon  = - \eta \zeta_L/3$, and for matter dominance ($a \propto \eta^2$) $\epsilon  = - \eta \zeta_L/5$. More generally, one can describe the transition from radiation to matter dominance using $a (\eta)= a_{\rm eq} (2 \alpha \eta + \alpha^2 \eta^2)$, where $\alpha$ is a constant which fixes the normalization of $\eta$, and $a_{\rm eq}$ is the value of the scale factor at radiation/matter equality. In this case eq.~\eqref{eps} gives
\be
\epsilon = \eta f(\eta) \zeta_L \;, \qquad  f(\eta) \equiv - \frac{20 + 15 \alpha \eta +3 \alpha^2 \eta^2}{15 (2 + \alpha\eta)^2} \;. \label{eps_MD}
\ee

Now we want to use the coordinate change above to study the second-order evolution of a short-wavelength mode $k_S$ under the modulation of the long mode $k_L$. 
In the squeezed limit, $k_S  \gg k_L$, the physical effect of the very long wavelength mode $\Phi_{L}$ simply amounts to a modification of the time and spatial coordinates of the short-wavelength mode \cite{Maldacena:2002vr,Creminelli:2004yq}. 
We start from a perturbed metric in the coordinates $\tilde \eta,\tilde x^i$, where the long mode has been reabsorbed in the background quantities,
\be
ds^2 = a^2 (\tilde \eta) \left[ - (1+  2 \tilde \Phi_S) d \tilde \eta^2 + (1-2 \tilde \Psi_S) \delta_{ij} d\tilde x^i d\tilde x^j \right] \;. \label{metric_2}
\ee
The potentials $\tilde \Phi_S$ and $\tilde \Psi_S$ can be written in terms of the coordinates $(\eta,x^i)$ using the coordinate transformation \eqref{ct}. They become (for this transformation in matter dominance see \cite{Fitzpatrick:2009ci})
\be
\tilde \Phi_S (\tilde \eta, \tilde x^i)= \tilde \Phi_S  +  \epsilon  \frac{\partial \tilde \Phi_S}{\partial  \eta} -\lambda x^i  \frac{\partial  \tilde\Phi_S}{\partial x^i}\;, \qquad 
\tilde \Psi_S (\tilde \eta, \tilde x^i) = \tilde \Psi_S  +  \epsilon  \frac{\partial \tilde \Psi_S}{\partial  \eta} -\lambda x^i  \frac{\partial  \tilde \Psi_S}{\partial x^i}\;. 
\ee
By performing the coordinate transformation \eqref{ct} on the entire metric \eqref{metric_2}, this can be written as a second-order perturbed one,
\be
ds^2 = a^2 (\eta) \left[ - e^{2 \Phi} d \eta^2 + e^{-2 \Psi} \delta_{ij} dx^i dx^j \right] \;, \label{metric_3}
\ee
with
\begin{align}
\Phi &= \Phi_S  + \Phi_L +  \epsilon  \frac{\partial \Phi_S}{\partial   \eta} +\zeta_L x^i  \frac{\partial  \Phi_S}{\partial x^i}\;, \label{Phi_trans} \\
\Psi & = \Psi_S  + \Psi_L +  \epsilon  \frac{\partial \Psi_S}{\partial  \eta} +\zeta_L x^i  \frac{\partial  \Psi_S}{\partial x^i}\;. \label{Psi_trans}
\end{align}
Here we have dropped the tildes to simplify the notation.
All the quantities labelled by a subindex $S$ are to be computed at first order only. Note that there are only scalar modes in the metric. Indeed, vector and tensor modes generated by second-order scalar perturbations are suppressed in the limit of $k_L \to 0$. Note also that we have used the exponential form for the metric \eqref{metric_3} to reabsorb cross-coupling terms between long and short modes in a nice form.

The coordinate transformation affects also other scalar quantities such as the temperature $T$, which transforms as
\be
T = T_S  + \epsilon  \frac{\partial T_S }{\partial  \eta}  + \zeta_L x^i \frac{\partial  T_S}{\partial x^i} \;. \label{T_trans}
\ee
Defining the temperature perturbation $\Theta$ as 
\be
T(\eta,x^i) \equiv \bar T (\eta) \left(1+ \Theta (\eta,x^i) \right) \;, 
\ee
one obtains, using $\bar T \propto 1/a$,\footnote{Note that the local term $- \HH \epsilon \Theta_S$ would have been absent, had we defined the temperature perturbation $\Theta$ in exponential form, i.e.~$T \propto \bar T e^\Theta$, such as for the potential $\Phi$ and $\Psi$.}
\be
\Theta  = \Theta_S  -  \HH \epsilon (1 + \Theta_S ) + \epsilon \frac{\partial \Theta_S}{\partial \eta}  +\zeta_L x^i \frac{\partial  \Theta_S}{\partial x^i}
\;. \label{Theta_trans}
\ee

For later purposes, it is convenient to give here also the effect of the coordinate redefinition on the fluid velocity. As the fluid four-velocity $u^\mu$ is a vector, under a coordinate change its components transform as
\be
u^\mu(\eta, x^i) = \frac{\partial x^\mu}{\partial \tilde x^\nu} u^\nu(\tilde \eta, \tilde x^i)\;.
\ee
From the expression above we obtain, for the three-velocity of the fluid $v^i$ ($u^i \equiv e^{\Psi} v^i/a$),
\be
v^i = v^i_S  + \epsilon \frac{\partial v^i_S}{\partial \eta}  +\zeta_L x^j \frac{\partial  v^i_S}{\partial x^j} \;. \label{V_trans}
\ee
In section~\ref{sec:comparison} we have checked numerically that these expressions accurately reproduce the second-order evolution in the squeezed limit.


\section{The effect of a long mode on the CMB anisotropies}
\label{sec:long}

In the limit of instantaneous recombination and considering only matter dominance, the observed CMB anisotropies $\Theta_{\rm obs}$ are given, in the squeezed limit, by
\be
\Theta_{\rm obs}(\n) = \left[ \Theta + \Phi + \n \cdot \vec v + \Phi \Theta +  \Phi^2/2 + (\Theta + \Phi) \n \cdot \vec v  \right] (\eta_{\rm rec},\vec x_{\rm rec})\;, \label{CMB}
\ee
where $\hat n$ is the direction of the line of sight. On the right-hand side, $\Theta$ is the intrinsic temperature fluctuations, $\Phi$ is the gravitational potential and $\vec v$ is the velocity of the baryon-photon fluid; these quantities are evaluated at the physical recombination time $\eta_{\rm rec}$  and position $\vec x_{\rm rec}$. A detailed derivation of this formula can be found in appendix~\ref{app:CMB} (see eq.~\eqref{CMB2}).

In order to study the effect of a long-wavelength mode on the CMB anisotropies, we will apply the change of coordinates \eqref{ct} on each of the terms on the right-hand side of this expression. Despite the assumptions in the derivation of eq.~\eqref{CMB}, i.e.~matter dominance and instantaneous recombination, the results that we will obtain are much more general. We will come back on this point at the end of this section.

In matter dominance eqs.~\eqref{Phi_trans}, \eqref{Theta_trans} and \eqref{V_trans} become
\begin{align}
\Phi & = \Phi_S + \Phi_L   +  \frac13 \Phi_{L} \frac{\partial \Phi_S}{\partial \ln \eta} -\frac53 \Phi_L  x^i  \frac{\partial \Phi_S}{\partial x^i} \;,  \label{Phi_t}\\
\Theta & = \Theta_S - \frac23 \Phi_{L} (1 + \Theta_S ) + \frac13  \Phi_{L} \frac{\partial \Theta_S}{\partial \ln \eta}  -\frac53 \Phi_{L} x^i  \frac{\partial \Theta_S}{\partial x^i} \;, \label{Theta_t} \\
v^j & = v^j_S  + \frac13 \Phi_L \frac{\partial  v^j_S}{\partial \ln \eta}  -\frac53 \Phi_{L} x^i  \frac{\partial v^j_S}{\partial x^i} \;, \label{v_t}
\end{align}
where we have used that in matter dominance $\zeta_L = -5 \Phi_L/3 $. 
We can now evaluate the effect of a long-wavelength mode on $\Theta_{\rm obs}$ by applying these equations to replace
$\Phi$, $\Theta$ and $\vec v$ on the right-hand side of eq.~\eqref{CMB}. 
The right-hand side rearranges nicely and the temperature anisotropy can be cast in terms of observable quantities, as
\be
\Theta_{\rm obs}(\n) =  \Theta_{{\rm obs},S} (\hat n) + \Theta_{{\rm obs},L}(\hat n) + \Theta_{{\rm obs},L}(\hat n) \left( 1  + \frac{\partial}{\partial \ln \eta_{\rm rec}}   - 5 \hat n \cdot \vec \nabla_{\hat n}  \right) \Theta_{{\rm obs},S} (\hat n) \;, \label{CMB_s}
\ee
with
\begin{align}
\Theta_{{\rm obs},S}(\hat n) & \equiv \left[ \Theta_S + \Phi_S + \n \cdot \vec v_S \right] (\eta_{\rm rec},\vec x_{\rm rec}) \;, \label{ThetaS} \\
\quad \Theta_{{\rm obs},L}(\hat n) & \equiv \frac13 \Phi_L (\eta_{\rm rec},\vec x_{\rm rec}) \;. \label{ThetaL}
\end{align}
In eq.~\eqref{CMB_s}, the dependence of $\Theta_{{\rm obs},S} ( \hat n)$ and $\Theta_{{\rm obs},L} ( \hat n)$ on $\eta_{\rm rec}$ can be read from eqs.~\eqref{ThetaS} and \eqref{ThetaL}.

In these expressions $\eta_{\rm rec}$ corresponds to the {\em physical} time of photon emission. At zeroth order $\eta_{\rm rec}$ is just given by the unperturbed recombination time $\eta_*$ and it is therefore space-independent.
In the presence of a long-wavelength mode,  recombination will not take place at the same coordinate time but at constant physical temperature. In matter dominance the coordinate transformation eq.~\eqref{ct} becomes
\be
\label{eq:MDredef}
\tilde \eta = \eta (1+ \frac13 \Phi_L)\;, \qquad \tilde x^i =  x^i(1 - \frac53 \Phi_L)\;.
\ee
Recombination takes place at constant $\tilde \eta=\eta_*$, which then
gives 
\be
\eta_{\rm rec} = \left(1-\frac{\Phi_L(\vec x)}{3} \right) \eta_*\;. \label{eta_rec}
\ee
Similarly, also the physical time of observation $\eta_{\rm obs}$, which at zeroth order coincides with the unperturbed observation time $\eta_0$, will be changed by the presence of a long-wavelength mode, $\eta_{\rm obs} = \left(1- {\Phi_L(\vec x)}/{3} \right) \eta_0$. 

The presence of the long-wavelength mode affects also the relation between the direction of observation $\hat n$ and the physical position at recombination $\vec x_{\rm rec}$. In the absence of $\Phi_L$, $\vec x_{\rm rec}$ is given by the zeroth-order geodesic equation, $\vec x_{\rm rec} = \hat n (\eta_0 - \eta_*) $, while in the presence of $\Phi_L$ it is given by
\be
\vec x_{\rm rec} (\hat n) = \hat n \left[ \left(1-\frac{\Phi_L(\vec x_0)}{3} \right) \eta_0 - \left(1-\frac{\Phi_L(\vec x_*)}{3} \right) \eta_* \right] + 2 \hat n \int_{\eta_*}^{\eta_0} \!  \Phi_L(\vec x) d \eta - 2  \int_{\eta_*}^{\eta_0} (\eta - \eta_*) \vec \nabla_\perp \Phi_L(\vec x) d \eta\;, \label{xrec_pert}
\ee
as a solution of the first-order geodesic equation. Note that in eq.~\eqref{CMB_s} we have rewritten $\vec x_{\rm rec} \cdot \vec \nabla_{\vec x_{\rm rec}}$ by using the zeroth-order geodesic equation, $\vec x_{\rm rec} = \hat n (\eta_0 - \eta_*) $.

Equation \eqref{CMB_s} satisfies an important consistency check \cite{Boubekeur:2009uk}: {\em a constant mode which is still far out of the Hubble radius  today does not affect the observed temperature anisotropy}. Let us see explicitly that when replacing $\Phi_L \equiv C$, where $C$ is a time and position independent constant, this constant drops out from the physical observables. To do so, one has to realize that there are hidden second-order contributions in the first two terms on the right-hand side of eq.~\eqref{CMB_s}. First of all, a constant mode also redefines the average temperature observed in the sky, i.e.
\be
\langle T (\hat n)\rangle = \langle \bar T (1 + \Theta_{{\rm obs},L}(\hat n)  ) \rangle=  \bar T \left( 1+ \frac{C}{3} \right) \;,
\ee
up to first order in $C$. The observed anisotropy must be defined in terms of this average temperature, see eq.~\eqref{Theta}, while equation \eqref{CMB_s} above is defined in terms of the unperturbed temperature $\bar T$. The variation of the average temperature changes the amplitude of small scale perturbations,
\be
\Theta_{{\rm obs},S} (\hat n) \cdot \frac{\bar T}{\langle T \rangle} = \Theta_{{\rm obs},S} (\hat n) \cdot \left(1 - \frac{C}{3}\right) \;.
\ee
This $C$-dependent contribution exactly cancels with the first term in parentheses of eq.~\eqref{CMB_s}, using $\Theta_{{\rm obs},L} = C/3$. For simplicity we did not include this effect of redefinition of the average temperature in eq.~\eqref{CMB_s}; indeed, this is only relevant when a mode is still out of the Hubble radius today, while modes of interest for the CMB bispectrum do not change the average temperature.

As discussed above, the presence of a long-wavelength mode perturbs the coordinate time of recombination, eq.~\eqref{eta_rec}. This introduces a second-order dependence on $C$ through eq.~\eqref{ThetaS},
\be
\label{eq:timerec}
\Theta_{{\rm obs},S}(\eta_{\rm rec},\hat n) = \Theta_{{\rm obs},S}(\eta_*,\hat n)- \frac{C}{3} \cdot \frac{\partial}{\partial \ln \eta_*} \Theta_{{\rm obs},S}(\eta_*,\hat n) \;,
\ee
at first order in $C$.
This cancels with the second term in parentheses in eq.~\eqref{CMB_s}, again using $\Theta_{{\rm obs},L} = C/3$.

The first-order relation between $\hat n$ and the position of the photon emission $\vec x_{\rm rec}$, eq.~\eqref{xrec_pert}, introduces another second-order contribution. For a constant mode, the lensing term in eq.~\eqref{xrec_pert} vanishes, while the first two terms yield
\be
\vec x_{\rm rec} (\hat n) = \hat n \left(1-\frac{C}{3} \right) (\eta_0 -  \eta_* ) + \hat n \;  2  C \; (\eta_0 - \eta_*) = \hat n \left(1+\frac{5 }{3} C\right) (\eta_0 -  \eta_* ) \;. \label{xrec_pert_C}
\ee
This $C$-dependent term cancels with the last term in parentheses in eq.~\eqref{CMB_s}. The constant mode redefines the coordinates; however, the relation between the direction of observation and the physical point of emission remains unchanged.

It is important to notice that the long-wavelength relation
\be
\label{eq:SWgeneral}
 \Theta_{{\rm obs},L} =  \frac13 \Phi_L = - \frac15 \zeta_L \;,
\ee
where $\Phi_L$ is the Newtonian potential during matter dominance, does not assume that recombination takes place in exact matter dominance. It holds also taking into account the transition from a radiation dominated Universe and including the effect of anisotropic stress. This is easy to understand as the standard calculations leading to the Sachs-Wolfe formula do not need to be done at recombination, but still hold later on when the Universe reaches exact matter dominance (assuming that the mode is still out of the Hubble radius). This can be checked explicitly as follows. The intrinsic temperature fluctuation for a mode which is out of the Hubble radius can be obtained by transforming to the unperturbed coordinate $\tilde \eta$. Given that the temperature scales as $T \propto 1/a$, it is given, independently of the background equation of state, by 
\be
\Theta = - \HH \epsilon \;.
\ee
Taking into account the gravitational redshift and the (early) ISW effect, we have
\be
\Theta_{\rm obs} = - \HH \epsilon + \Phi_{\rm rec} + \int_{\eta_{\rm rec}}^{\eta_{\rm MD}} d \eta \; (\Phi' + \Psi')= - \HH \epsilon - \Psi_{\rm rec} + 2 \Phi_{\rm MD} = \zeta + 2 \Phi_{\rm MD} = - \frac15 \zeta \;,
\ee
where we have used that $\Phi_{\rm MD} = \Psi_{\rm MD}$, eqs.~\eqref{conds} and \eqref{eq:lambdazeta} and the matter-dominance relation $\zeta = -5 \Phi/3 $. This implies that, though our derivation of eq.~\eqref{CMB_s} holds in matter dominance, the first and third terms in parentheses---which are the ones that enter in the final bispectrum---will remain the same also when we depart from exact matter dominance. Indeed, eq.~\eqref{eq:SWgeneral} implies that the relationship between the long-wavelength temperature fluctuation and the spatial redefinition of the coordinates, i.e.~$\zeta$, remains the same, and this fixes the factor $-5$ in \eqref{CMB_s}. Moreover, the factor of $1$ in parentheses remains the same. It just depends on the change in the average temperature given by the long mode and it is required by the validity of the previous argument about the cancellation of an homogeneous mode.

Another approximation that we have made so far, which can be relaxed without affecting the final result, is that temperature perturbations for the short modes are created instantaneously at recombination. Of course recombination is not instantaneous and moreover perturbations are also generated by the early ISW effect. However, it is easy to see that our results are not changed, because the long mode is taken to be out of the Hubble radius during the whole period when recombination happens and matter dominance is finally reached. For example, the fact that recombination is not instantaneous is encoded in the presence of a window function $g$ which weights the source term $S$, schematically
\be
\int d \eta \; g(\eta) S(\eta) \;.
\ee
Now the long mode will modify both the source term and the window function, through a time redefinition: the perturbed source term will be equal to the unperturbed one once written in the tilded coordinates $S(\eta) = S_0(\tilde \eta)$ and the same will hold for the window function (which has a different transformation property, as it is a probability density in $\eta$): $g(\eta) d\eta = g_0 (\tilde \eta) d\tilde \eta$. The integral remains the same and this parallels the cancellation discussed after eq.~\eqref{eq:timerec} in the instantaneous limit. The only effect of the time redefinition is the geometrical effect of equation \eqref{xrec_pert}. Along the same lines, one can show that also the Silk damping effect is taken into account.


\section{The bispectrum in the squeezed limit}
\label{sec:bispectrum}
In this section we derive the  bispectrum in the squeezed limit, eq.~\eqref{formula1}, starting from equations \eqref{CMB_s} and \eqref{xrec_pert}. The idea is to calculate how a long mode affects the short-scale 2-point function and from this to get to the bispectrum, similarly to what done in the derivation of the consistency relation for the primordial 3-point function \cite{Maldacena:2002vr,Creminelli:2004yq,Cheung:2007sv}. For simplicity we work in the flat-sky approximation where the angular direction can be written as $\hat n = (1, \vec m)$ and the multipole expressions are simply Fourier transforms with respect to the $\vec m$'s. 

Let us examine eq.~\eqref{xrec_pert}. The last term gives lensing and we are going to discuss it separately below. The first term in brackets does not contribute to the 3-point function as it depends only on the Newtonian potential at observation. The second term in brackets changes the distance to the last scattering surface, but its effect is suppressed by $\eta_*/\eta_0$ and can thus be neglected. The second term in eq.~\eqref{xrec_pert}, which involves the integral of $\Phi$ along the line of sight, is the so-called Shapiro time delay \cite{Hu:2001yq}. This is suppressed because the integral tends to average out along the line of sight, unless the mode is of the order of the present Hubble scale: this effect can be neglected if we disregard the very first multipoles. In eq.~\eqref{CMB_s} the time derivative combines with the change in the recombination time, eq.~\eqref{eta_rec}, as we discussed in the previous section: one is left with the geometrical effect that light is emitted at a different coordinate time, but this is again suppressed by $\eta_*/\eta_0$. The same geometrical suppression affects the radial part of the spatial derivative. Therefore, the effect of the long mode on the short-scale 2-point function only amounts to a stretching perpendicular to the line of sight and to the constant term, the first in parentheses in eq.~\eqref{CMB_s}. This gives
\be
\langle \Theta_{\rm obs}(\vec m_2) \Theta_{\rm obs}(\vec m_3) \rangle_L = \langle \Theta_{\rm obs}(\vec m_2) \Theta_{\rm obs}(\vec m_3) \rangle_0 + \Theta_{{\rm obs},L}(\vec m_1) \left( 2   - 5 \vec m \cdot \vec \nabla_{m}  \right) \langle \Theta_{{\rm obs}} (\vec m_{2}) \Theta_{{\rm obs}} (\vec m_3)\rangle\;, 
\ee
where the long-wavelength mode is evaluated at the midpoint $\vec  m_1 =  (\vec m_2 + \vec m_3)/2$. Notice that the short scale 2-point function only depends on the distance $\vec m = \vec m_3 - \vec m_2$. The factor of $2$ in the parenthesis takes into account that the long mode affects both $\Theta_{\rm obs}(\vec m_2)$ and $\Theta_{\rm obs}(\vec m_3)$.

To get to the 3-point function one has to correlate the expression above with the long-wavelength temperature and then take the Fourier transform. These steps are the same as in the derivation of the 3-point function consistency relation \cite{Maldacena:2002vr,Creminelli:2004yq,Cheung:2007sv} and one gets 
\be
B_{l_L l_S l_S} = C_{l_L} C_{l_S} \left(2 + 5 \frac{d \ln (l_S^2 C_{l_S})}{d \ln l_S}\right) \;.  \label{final_conv}
\ee
Note that the long-mode power spectrum $C_{l_L}$ should be understood only to contain the Sachs-Wolfe effect and not the late ISW. The same disclaimer applies to the analogous equations below.

To this we have to add the contribution due to lensing, which comes from the last term of eq.~\eqref{xrec_pert}. In this paper we are not dealing with late-time effects, but only with effects which occur close to the surface of last scattering.  This means that we only have the contribution coming from the correlation of the lensing potential due to the long mode with the temperature anisotropy created at last scattering, and not the (much bigger) effect due to the correlation between lensing and (late) ISW effect \cite{Seljak:1998nu,Goldberg:1999xm,Hanson:2009kg,Lewis:2011fk}. In the limit in which the lensing mode is out of the Hubble radius at recombination, the lensing bispectrum is easy to calculate and is given by \cite{Boubekeur:2009uk}
\be
B_{l_L l_S l_S} =  6 C_{l_L} C_{l_S} \left[2 \cos 2 \theta   - (1+\cos 2 \theta) \frac{d \ln (l_S^2 C_{l_S})}{d \ln l_S} \right]\;, \label{final_lensing}
\ee
where $\theta$ is the angle between the long and short modes.
It is important to stress that, although this result was obtained assuming perfect matter dominance and therefore neglecting the role of the early ISW and the anisotropic stress, the final result does not change when these effects are included, similarly to what happens with eq.~\eqref{eq:SWgeneral}. Indeed the correlation between temperature and lensing potential peaks at the time when the long mode comes back into the Hubble radius (before that it is suppressed by the lensing window, later it loses correlation with the temperature fluctuation \cite{Boubekeur:2009uk}). This implies that \eqref{final_lensing} is not modified by recombination taking place when radiation is not completely negligible, because sufficiently long modes come back into the horizon in exact matter dominance and the relationship between the Newtonian potential and the temperature is unchanged, see eq.~\eqref{eq:SWgeneral}.

The two equations \eqref{final_conv} and \eqref{final_lensing} have a similar structure and can be put together to give
\be
B_{l_L l_S l_S}=  C_{l_L} C_{l_S} (1+  6 \cos 2 \theta)   \left( 2 - \frac{d \ln (l_S^2 C_{l_S})}{d \ln l_S}  \right)  
 \label{final_formula}\;,
\ee
which is the main theoretical conclusion of the paper. In the squeezed limit, a local primordial non-Gaussianity would give a bispectrum of the form $- 12 f^{\rm loc}_{\rm NL} C_{l_L} C_{l_S}$. Therefore, this bispectrum corresponds to 
\be
f^{\rm loc}_{\rm NL} = -\frac16 (1+  6 \cos 2 \theta) \left(1- \frac 12 \frac{d \ln (l_S^2 C_{l_S})}{d \ln l_S} \right)   \;. \label{fNL_final}
\ee
Note that for a scale invariant $C_{l_S}$, which occurs when also the short modes are out of the Hubble radius at recombination, the second term in parentheses cancels and we recover the bispectrum computed in \cite{Boubekeur:2009uk} in the squeezed limit.

In fig.~\ref{fig:fnleq} we plot the angle independent part of eq.~\eqref{fNL_final}, i.e.~when we neglect the $6 \cos 2\theta$.  We see that the final result is quite small over the whole range of $l_S$:  $f^{\rm loc}_{\rm NL} \aplt 1$. The $\theta$-dependent part is larger by a factor of $6$, but it will not contaminate any primordial signal, as it averages to zero when summed over all relative orientations of the modes.\footnote{All primordial shapes, to our knowledge, have an $f_{\rm NL}^{\rm loc}$ which is independent on the relative orientation of the modes in the squeezed limit.}
\begin{figure}[t]
\begin{center}
\includegraphics[scale=1]{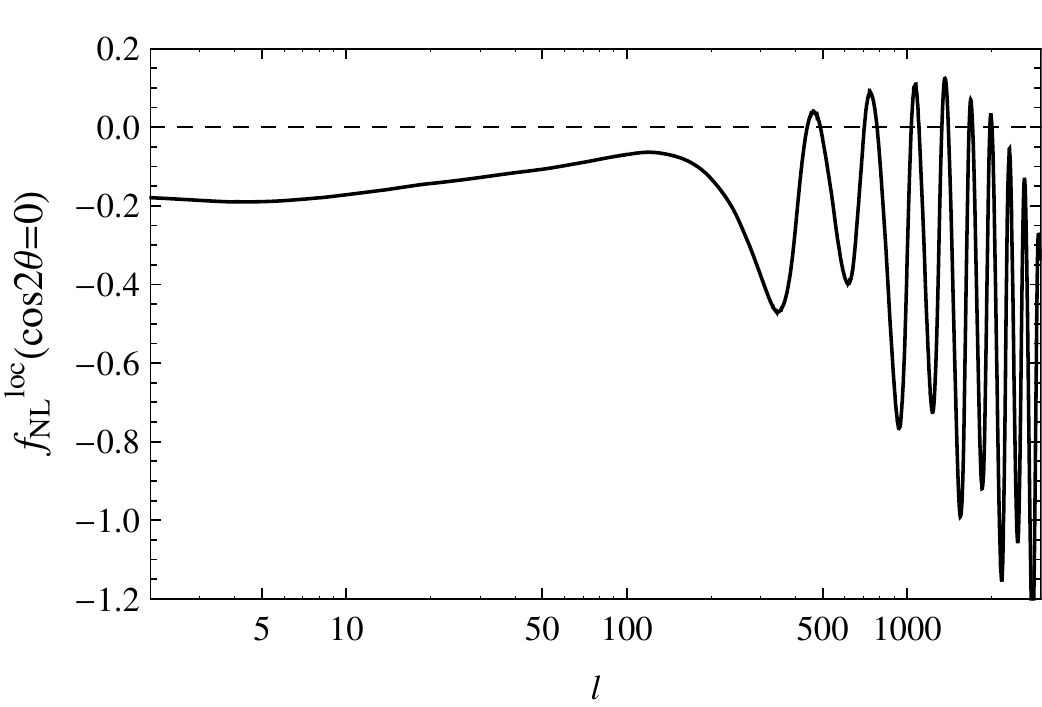}  
\end{center}
\caption{\small Plot of $-1/6 \cdot \left(1- 1/2 \cdot d \ln (l_S^2 C_{l_S})/d \ln l_S\right)$, which acts as an effective $f^{\rm loc}_{\rm NL}$ in our final formula \eqref{final_formula}.}
\label{fig:fnleq}
\end{figure}

Notice that there is a partial cancellation between the effect coming from the redefinition of the spatial coordinates and the isotropic part of lensing, as already noticed in \cite{Creminelli:2004pv}. This is easy to understand as overdense regions around the surface of last scattering will have a positive convergence, and thus move the short-scale 2-point function towards larger angular scales; at the same time these regions will have $\Phi <0$ and this, through eq.~\eqref{eq:MDredef}, moves the 2-point function in the opposite direction.\footnote{In \cite{Creminelli:2004pv} a wrong overall minus sign in front of the two effects was introduced, when moving from real to Fourier space. This does not affect the partial cancellation between the two contributions. } This cancellation considerably suppresses the final result, so that it would be quite misleading to separate lensing from the other effects. Notice that the effect of lensing close to the last scattering surface is included (though dwarfed by the much larger ISW-lensing signal) in the studies of lensing-induced bispectrum \cite{Seljak:1998nu,Goldberg:1999xm,Hanson:2009kg,Lewis:2011fk}; therefore, one must be careful not to double-count it. 

The $\theta$-dependent term of eq.~\eqref{final_formula}, which gives zero when we average over the relative orientation of the long and short modes,  is entirely due to the shear (i.e.~the traceless) part of lensing. This term gives zero when the short-scale 2-point function has $C_{l_S} =$ const, i.e.~for  a white-noise spectrum, with a real-space correlation which is a delta function. This is easy to understand as shear just remaps the points, preserving the area, so that a white noise remains such (see for example \cite{Lewis:2011fk}). Also the first term in eq.~\eqref{final_formula}, which includes the isotropic part of lensing and equation \eqref{final_conv}, cancels for a white spectrum although we did not find any simple explanation for that.\footnote{Notice that eq.~\eqref{final_lensing} holds only for a scale-invariant spectrum of the long mode. Indeed, when we deviate from scale-invariance, the relationship between the long-mode spectrum $C_{l_L}$ and the correlation $C^{T \psi}_{l_L}$ between temperature fluctuation and lensing potential changes. In flat sky the temperature power spectrum in the Sachs-Wolfe limit is given by 
\be
C_{l_L}= \frac{1}{3 D^2} \cdot \int_{-\infty}^{+\infty} d k_\perp \frac{A}{(k_\perp^2 + l_L^2/D^2)^{(3-(n_s-1))/2}} = A \cdot \frac{2 \sqrt{\pi } \;\Gamma \left(\frac{3}{2}-\frac{n_s}{2}\right)
   }{3 \Gamma
   \left(2-\frac{n_s}{2}\right)} \cdot \frac{D^{1-n_s}}{l_L^{3-n_s}} \;,
\ee
where $A$ is the normalization of the power spectrum, $D \equiv \eta_0 - \eta_*$ is the distance from the last-scattering surface and $n_s-1$ is the deviation from a scale-invariant spectrum. In the same limit the correlation between temperature and lensing potential is given by 
\be
C^{T \psi}_{l_L}= -\frac{2}{3D^2} \cdot \int_{-\infty}^{+\infty} d k_\perp \int_{\eta_*}^{\eta_0} d \eta \frac{\eta-\eta_*}{D^2} \frac{A}{(k_\perp^2 + l_L^2/D^2)^{(3-(n_s-1))/2} } e^{i k_\perp (\eta -\eta_*)}  = - A\frac{4 \sqrt{\pi } \;\Gamma \left(\frac{5}{2}-\frac{n_s}{2}\right)
   }{3 \Gamma
   \left(2-\frac{n_s}{2}\right)}  \cdot \frac{D^{1-n_s}}{l_L^{5-n_s}} \;.
\ee
We see that the relative coefficient between the two depends on the tilt
\be
\frac{C^{T \psi}_{l_L}}{C_{l_L}} = -\frac{3-n_s}{l_L^2} \;.
\ee
Thus, the overall coefficient of eq.~\eqref{final_lensing} changes when departing from scale-invariance on large scales, while eq.~\eqref{final_conv} does not. This, in particular, means that the cancellation between the two effects for a white-noise spectrum on short scales, $C_{l_{S}} =$ const, only occurs for a scale-invariant large-scale spectrum.}

To end this section, let us discuss an important point of eqs.~\eqref{final_conv}, \eqref{final_lensing} and \eqref{final_formula}. As we said many times, the effect of a super-Hubble mode on the short ones simply amounts to a coordinate redefinition of the background. Note that the long-wavelength modes $k_L$ that we are considering are already super-horizon when the short modes are quantum mechanically generated during inflation and they remain so until after recombination. Thus, also the primordial 2-point function of the short modes is rescaled by the presence of the long-wavelength modes. This implies that our results, eqs.~\eqref{final_conv}, \eqref{final_lensing} and \eqref{final_formula}, already take into account the single-field Maldacena consistency relation \cite{Maldacena:2002vr,Creminelli:2004yq,Cheung:2007sv}, which relates the primordial 3-point function in the squeezed limit to the spectral tilt. Indeed, the tilt of the primordial spectrum contributes to the scale dependence of the temperature 2-point function above. In other words, {\em our results already include the contribution from the primordial non-Gaussianity in any single-field model.} Indeed, we checked that this is the case also when a substantial departure from scale-invariance in the primordial spectrum is allowed. Multi-field models of inflation can produce additional non-Gaussianity in the squeezed limit (see for instance \cite{Lyth:2001nq,Dvali:2003ar,Vernizzi:2006ve}) and this should be added to our results.


\section{Comparison with numerical calculations}
\label{sec:comparison}

Before comparing the results of the previous section with CMBquick, we would like to stress that the coordinate transformation derived in section~\ref{sec:coord} can be used to reproduce the evolution of perturbations at  second order in the squeezed limit. In particular, eqs.~\eqref{Phi_trans}, \eqref{Psi_trans}, \eqref{Theta_trans} and \eqref{V_trans}
give the same evolution as that obtained from Einstein's equations when one of the two modes in the second-order source terms remains out of the Hubble radius. 

As an example, we consider the radiation to matter transition where $\epsilon$ is given by eq.~\eqref{eps_MD}, and we study the second-order evolution of a small-scale mode of the potential $\Phi$ and of the radiation energy density perturbation $\delta_{\rm rad}$ in the perfect fluid approximation.
\begin{figure}[t]
\begin{center}
{\includegraphics[scale=0.74]{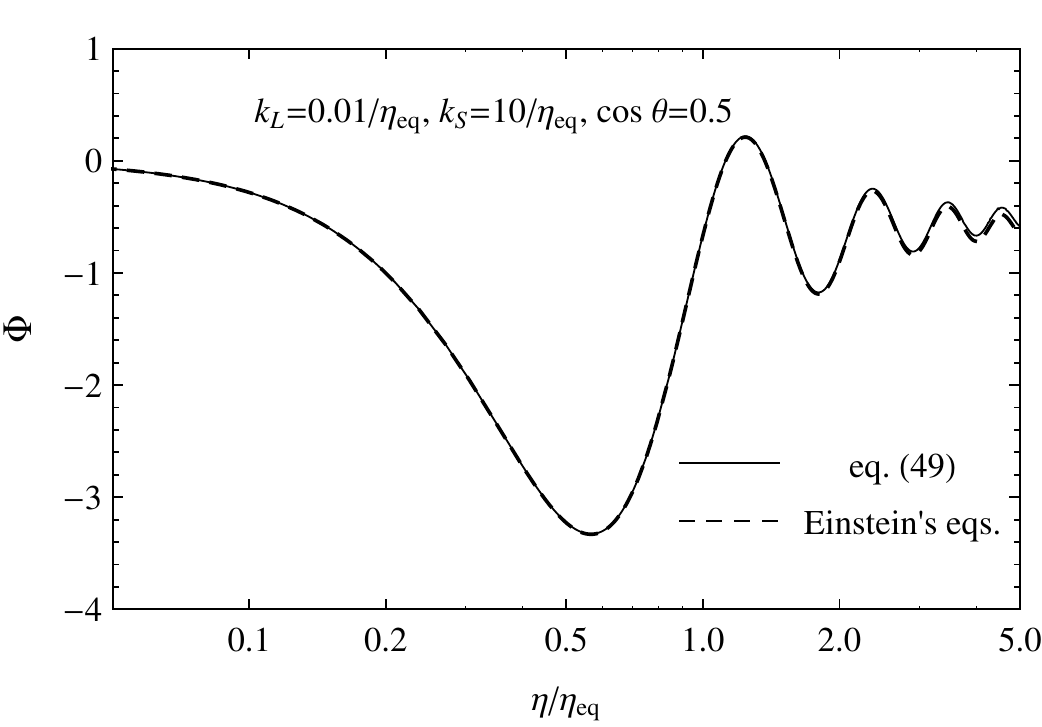}}
{\includegraphics[scale=0.76]{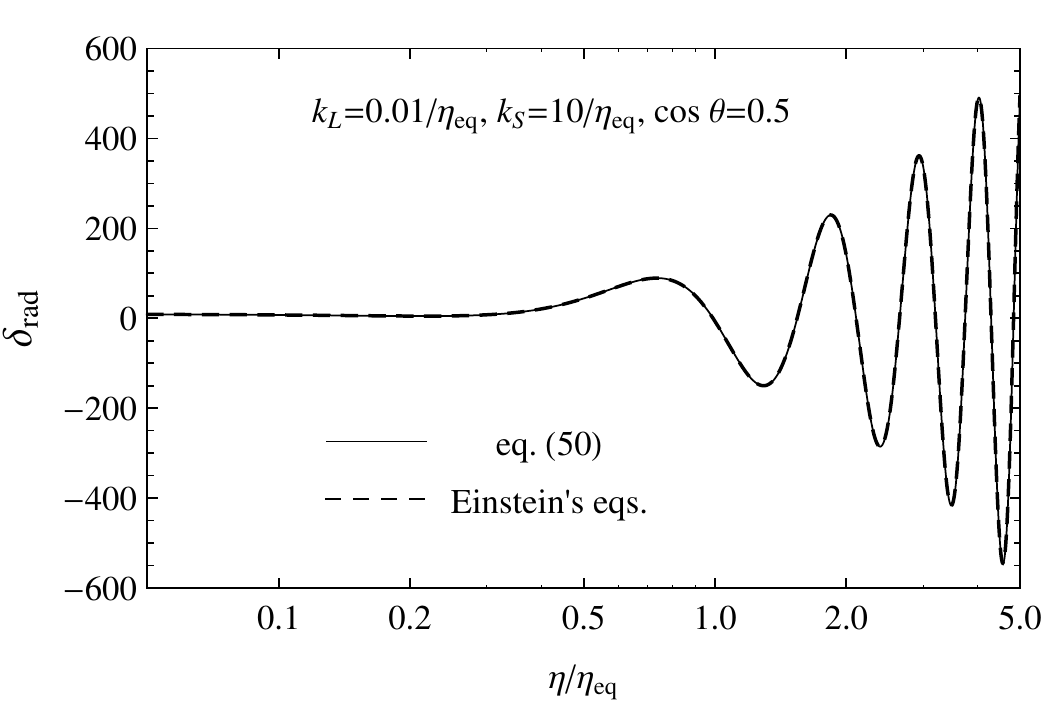}}
\caption{Comparison of the effect of the coordinate transformation \eqref{ct} on the gravitational potential $\Phi$ and on the radiation energy density $\delta_{\rm rad} = 4 \Theta$ to the full second-order evolution in the squeezed limit. We consider the radiation-to-matter transition in the fluid approximation. On the left-hand panel we compare eq.~\eqref{Phi_sec} (solid line) to the numerical solution of Einstein's equation (dashed line). On the right-hand panel we do the same for eq.~\eqref{delta_sec}. We chose $k_L = 0.01 \eta^{-1}_{\rm eq}$, $k_S = 10 \eta^{-1}_{\rm eq}$ and $\cos \theta=0.5$.
}
\label{fig:comparison1}
\end{center}
\end{figure}
Fourier transforming eqs.~\eqref{Phi_trans} and \eqref{Theta_trans} (we use $\Theta = \delta_{\rm rad}/4$) and using eq.~\eqref{eps_MD}  yields
\begin{align}
T_\Phi^{(2)} (k_S,\eta) &=  f(\eta) \frac{\partial \; T_\Phi^{(1)}(k_S,\eta)}{\partial \ln \eta} - \frac{\partial \; T_\Phi^{(1)}(k_S,\eta)}{\partial \ln k_S}\;, \label{Phi_sec}\\
T_{\delta_{\rm rad}}^{(2)} (k_S,\eta) &=  - 4 f(\eta) \eta \HH  \; T^{(1)}_{\delta_{\rm rad}}(k_S, \eta) + f(\eta) \frac{\partial \; T_{\delta_{\rm rad}}^{(1)}(k_S,\eta)}{\partial \ln \eta} - \frac{\partial \; T_{\delta_{\rm rad}}^{(1)}(k_S,\eta)}{\partial \ln k_S} \;, \label{delta_sec}
\end{align}
where $f(\eta)$ can be read from eq.~\eqref{eps_MD} and by $T^{(1)}$ and $T^{(2)}$ we denote the first and second-order transfer functions. For a generic perturbation $X$ these are defined by 
\begin{align}
X^{(1)}_{\vec k} (\eta) & \equiv T_X^{(1)}(k,\eta) \zeta_{\vec k}\;, \\
X^{(2)}_{\vec k} (\eta) & \equiv \int \frac{d^3 k_L\; d^3 k_S}{(2 \pi)^3}   \delta(\vec k - \vec k_L -\vec k_S)\, T_X^{(2)}(k,k_L,k_S;\eta)\, \zeta_{\vec k_L} \zeta_{\vec k_S}  \;.
\end{align}
In fig.~\ref{fig:comparison1} we compare eqs.~\eqref{Phi_sec} and \eqref{delta_sec} with the solution of Einstein's equations at second order (see, for instance, \cite{Fitzpatrick:2009ci}). We chose $k_L = 0.01 \eta^{-1}_{\rm eq}$, $k_S = 10 \eta^{-1}_{\rm eq}$ and $\cos \theta=0.5$, where $\theta$ is the angle between the long and the short mode and $\eta_{\rm eq}$ is the time of radiation-matter equality. The agreement is very good and gets better as we increase the ratio $k_S/k_L$.

Let us go back to the bispectrum.
As discussed before, in the squeezed limit our results can be also used as a consistency check of numerical Boltzmann codes at second order.
Here we compare the bispectrum computed in the previous section with that produced by CMBquick \cite{cyril}.

Since the lensing bispectrum is a well-known result in the literature, we have chosen to remove the lensing effect from the code. This is easy to do, because the code parallels our calculation using Newtonian gauge. In this way we do not have any $\theta$-dependence in the bispectrum and we avoid numerical cancellations between the isotropic part of lensing and the space redefinition effect. Thus, we  concentrate on reproducing the bispectrum given by eq.~\eqref{final_conv}. 
The comparison is direct because CMBquick computes the bispectrum using the flat-sky approximation.

\begin{figure}[t]
\begin{center}
{\includegraphics[scale=0.75]{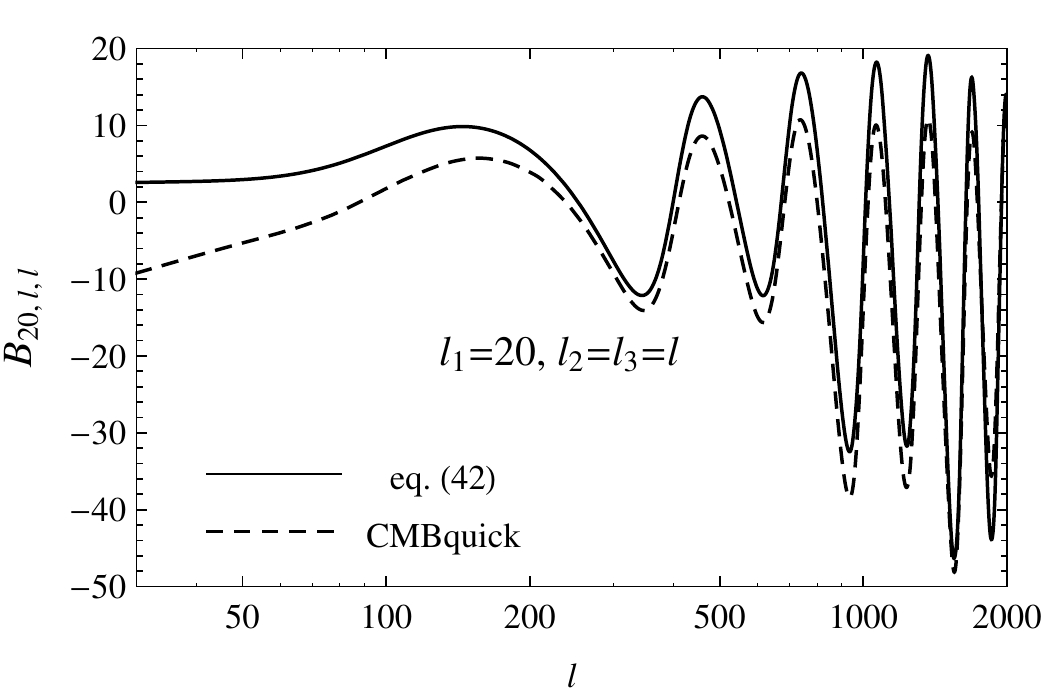}}
{\includegraphics[scale=0.75]{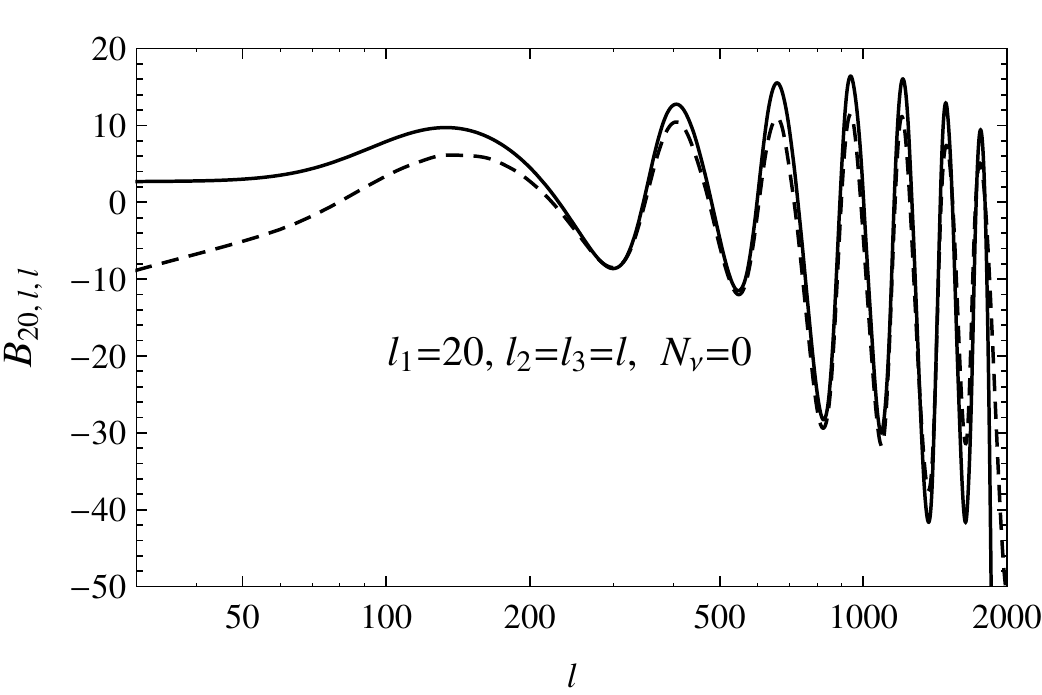}}
{\includegraphics[scale=0.75]{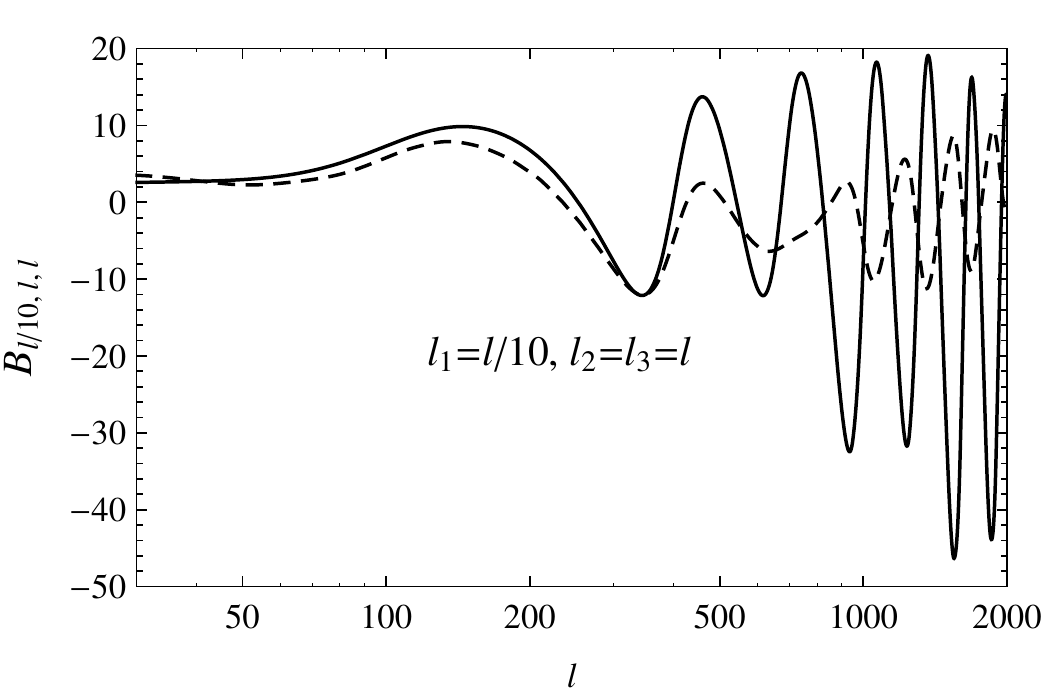}}
{\includegraphics[scale=0.75]{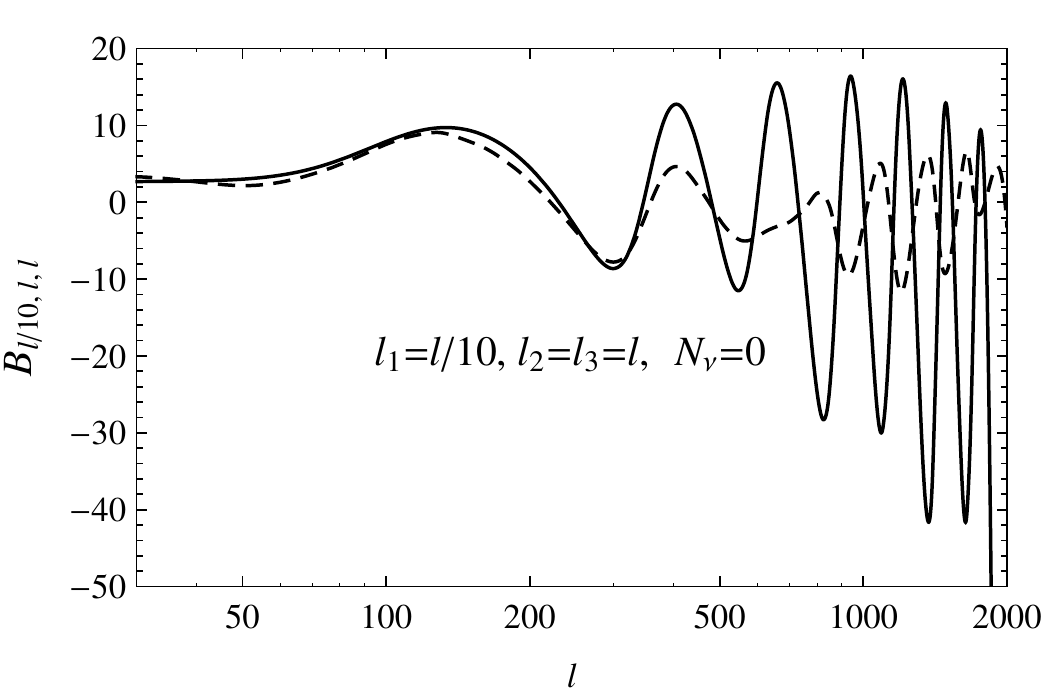}}
\caption{Comparison of eq.~\eqref{final_conv} with the bispectrum in the flat-sky approximation produced by CMBquick in the absence of lensing. Cosmological parameters are the best-fit ones of WMAP5 \cite{Komatsu:2008hk}. Solid lines represent our equation while dashed lines the bispectrum computed by the code. In the top panels $l_1=20$ and $ l_2=l_3=l$; in the bottom panels $l_1=l/10$ and $ l_2=l_3=l$. For the top panels the squeezed limit corresponds to $l \apgt 300$ and for the bottom panels to $l \aplt 300$. On the right-hand panels we have removed neutrinos and set the photon temperature today to $T=3.08$ K to have the same radiation density.}
\label{fig:comparison2}
\end{center}
\end{figure}
In fig.~\ref{fig:comparison2} we compare the bispectrum from CMBquick with eq.~\eqref{final_conv}.
Solid lines represent our equation while dashed lines the bispectrum computed by the code. 
In all plots we have used the best-fit cosmological parameters of WMAP5 \cite{Komatsu:2008hk}. In the top panels we have fixed one of the multipoles to be $l_1=20$ and we have plotted the bispectra as a function of the other two, taken to be equal $l_2=l_3=l$. We expect the two curves to converge in the limit $l \gg 20$.
To help the comparison at low multipoles, in the bottom panels we have taken $l_1 = l/10$ and $l_2=l_3=l$. Here we expect the two curves to converge for small $l$.
The code agrees reasonably well with our formula, for $l\apgt 200$ in the top panels and for $l \aplt 300$ in the bottom panels. The agreement improves when we remove neutrinos, as shown on the right-hand panels. In this case, in order to maintain the same amount of radiation as in the left-hand panels, we have increased the photon temperature today to $T=3.08$ K.
The reason for this improvement is interely numerical. Neutrinos require finer integration steps and by removing them we can reach a better accuracy for the same computing time. 
There is another important source of inaccuracy in the results of the code. As the second-order numerical integration is particularly slow, we have integrated the second-order sources along the line of sight only in a finite range of conformal time, i.e.~between $\eta = 230 $ Mpc to $\eta= 350$ Mpc (the peak of the visibility function being at $\sim 280$ Mpc), neglecting part of the early ISW effect. In conclusion, the code is reliable in the regime that we studied and we are confident that the residual discrepancies can be removed by a proper optimization. Note that, in comparing the code with our analytical approach, we corrected some numerical inaccuracies which may affect the results of the current version of \cite{Pitrou:2010sn}.


\section{Contamination and observability}
\label{sec:contamination}

In the flat-sky approximation, the estimator of a local primordial signal is given by \cite{Komatsu:2003iq,Hu:2000ee}
\be
{\cal E}_{\rm loc} = \frac1{\cal N}\int \frac{d^2 l_1}{(2 \pi)^2} \frac{d^2 l_2}{(2 \pi)^2} \frac{d^2 l_3}{(2 \pi)^2} (2 \pi)^2 \delta(\vec l_1 +\vec l_2 +\vec l_3) \frac{B_{\rm loc}(l_1,l_2,l_3)}{6 C_{l_1} C_{l_2}C_{l_3}} a(\vec l_1) a(\vec l_2)a(\vec l_3) \;, 
\ee
where $B_{\rm loc}(l_1,l_2,l_3)$ is the bispectrum for local non-Gaussianities when $f_{\rm NL}^{\rm loc}=1$ and
the normalization factor that makes the estimator unbiased ($\langle {\cal E}_{\rm loc} \rangle = f_{\rm NL}^{\rm loc}$) is 
\be
{\cal N} = \frac{1}{\pi} \int  \frac{d^2 l_2 \, d^2 l_3}{(2 \pi)^2} \frac{[B_{\rm loc}(l_1,l_2,l_3)]^2}{6 C_{l_1} C_{l_2}C_{l_3}} \;.
\ee
Thus, if the local estimator above is applied to an arbitrary signal $B_X (l_1,l_2,l_3)$, this will contaminate the measurement by a value of $f_{\rm NL}^{\rm loc}$ given by
\be
f_{\rm NL}^{\rm loc} = \frac1{\cal N} \cdot \frac{1}{\pi} \int  \frac{d^2 l_2 \, d^2 l_3}{(2 \pi)^2} \frac{B_{\rm loc}(l_1,l_2,l_3) B_X (l_1,l_2,l_3)}{6 C_{l_1} C_{l_2}C_{l_3}} \;. \label{contamination}
\ee

Let us apply eq.~\eqref{contamination} to our bispectrum \eqref{final_formula} to compute its contamination on a measurement of a local signal.  
We evaluate the integrals\footnote{In order to solve the integrals in eq.~\eqref{contamination} it is convenient to use that
\be
 \int d^2 l_2 \, d^2 l_3 \, F(l_1,l_2,l_3) =  4 \pi \int \frac{ l_1 dl_1\, l_2 dl_2\,l_3 dl_3}{(-l_1^4 -l_2^4 -l_3^4 + 2 l_1^2 l_2^2 + 2 l_2^2 l_3^2 + 2 l_1^2 l_3^2)^{1/2}} \, F(l_1,l_2,l_3)\;,
\ee
where $F$ is a function of the moduli of the multipoles.}
 in the squeezed limit and take into account the triangular inequalities by choosing $2 \le l_1 \le 100$, $20\, l_1\le l_2 \le l_{\rm max}$ and $l_2-l_1 \le l_3 \le {\rm min}(l_1+l_2,l_{\rm max})$. We find 
 \be
 f_{\rm NL}^{\rm loc} = -0.39\;, 
 \ee
 for $l_{\rm max} = 2000$. This value is visually confirmed by fig.~\ref{fig:fnleq}, where, for $l \sim 2000$, the plotted equivalent $f_{\rm NL}^{\rm loc}$ oscillates around $\sim -0.4$.  Note that the anisotropic part of lensing, proportional to $\cos 2 \theta$, does not contribute to this value because the integration over the angle averages to zero. However, the cancellation due to the isotropic part of lensing is crucial to obtain such a small result.\footnote{For $l_{\rm max} =2000$, the contamination from the bispectrum in eq.~\eqref{final_conv} is $f_{\rm NL}^{\rm loc} = 0.94$, while that from lensing alone, eq.~\eqref{final_lensing}, is $f_{\rm NL}^{\rm loc} = -1.33$.}
As the contamination is given by the ratio between two integrals, this result is quite solid and changes little when we vary the range of integration keeping $l_{\rm max}$ fixed. 
We conclude that the bias introduced by this effect on a Planck search for a primordial non-Gaussian signal will be negligible. 
 The contamination increases if we take higher values of $l_{\rm max}$, as confirmed by the behavior of $f_{\rm NL}^{\rm loc}$ in fig.~\ref{fig:fnleq}. For instance, $f_{\rm NL}^{\rm loc} = -0.48$ for a futuristic experiment with $l_{\rm max}= 3000$.

\begin{figure}[!!!h]
\begin{center}
{\includegraphics[scale=1]{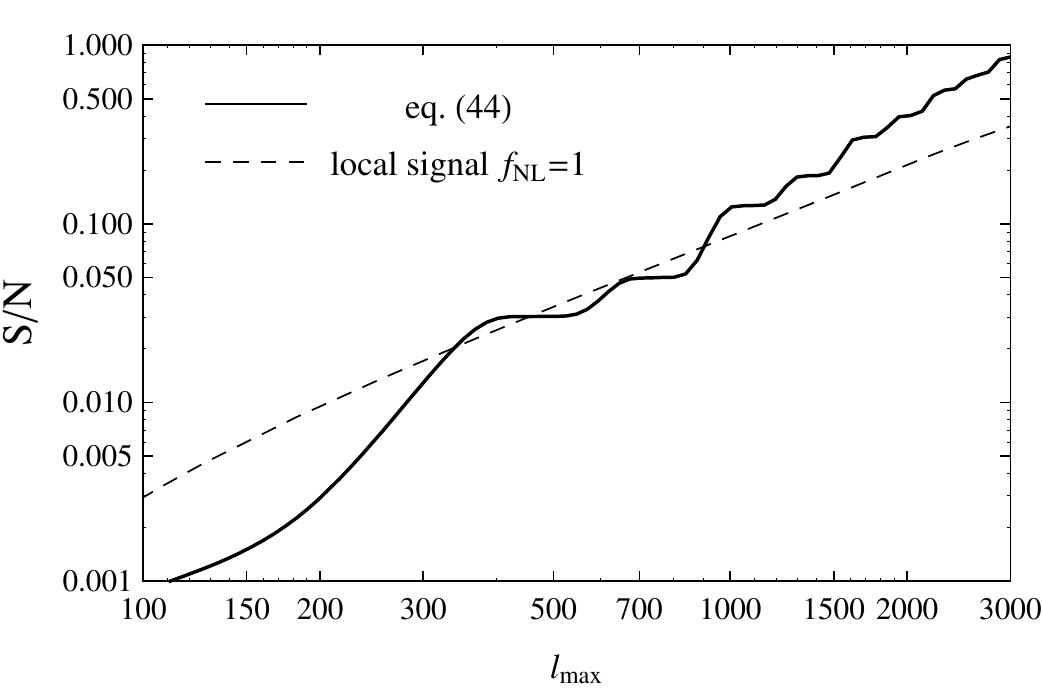}}
\caption{Signal-to-noise ratio for the bispectrum in eq.~\eqref{final_formula}, evaluated summing over squeezed configurations only, as a function of $l_{\rm max}$. This roughly oscillates around $\sim 10^{-7} l_{\rm max}^2$. For comparison, the dashed line shows $S/N$ for a primordial local signal with $f_{\rm NL}^{\rm loc}=1$. }
\label{fig:sn}
\end{center}
\end{figure}
Let us study now the observability of the bispectrum in eq.~\eqref{final_formula}. The possibility of measuring a bispectrum depends on its signal-to-noise ratio $S/N$. In the flat-sky approximation this is given by \cite{Hu:2000ee}
\be
\left(\frac{S}{N} \right)^2 = \frac{1}{\pi} \int  \frac{d^2 l_2 \, d^2 l_3}{(2 \pi)^2} \frac{[B (l_1,l_2,l_3)]^2}{6 C_{l_1} C_{l_2}C_{l_3}} \;.
\ee
Plugging eq.~\eqref{final_formula} in this expression and considering all cyclic permutations we obtain the signal-to-noise shown in fig.~\ref{fig:sn} (solid line) as a function of the maximum multipole $l_{\rm max}$. For comparison, we have also plotted the signal-to-noise for a primordial local signal with $f_{\rm NL}^{\rm loc}=1$ (dashed line). The signal-to-noise from eq.~\eqref{final_formula} grows approximately as $l_{\rm max}^2$ and not as $l_{\rm max}$ as one would naively expect by simply counting the number of available modes. Indeed, the effect of the exponential damping at high-$l$ contributes to the logarithmic derivative in eq.~\eqref{fNL_final} with a term which increases roughly linearly in $l_{\rm max}$. Note that, in the estimate of the signal-to-noise, the anisotropic part of lensing does not cancel and gives a large contribution to the signal.

From fig.~\ref{fig:sn} we see that this effect is not measurable by Planck. Note however that we have integrated only over configurations in the particular squeezed limit in which one of the modes is out of the Hubble radius, i.e.~for $l_L \le 100$. Thus, this $S/N$  represents only a fraction of what one would get by summing over all configurations of  the full bispectrum generated by a complete numerical Boltzmann code.
The fact that the contamination from second-order perturbations will not be relevant for Planck is confirmed by other studies \cite{Khatri:2008kb,Senatore:2008wk,Nitta:2009jp},  which focused on particular second-order effects.

Note that even though eq.~\eqref{final_formula} includes the primordial non-Gaussianity in the squeezed limit for single-field models, for the tiny value of the spectral index that we consider ($n_s-1 = 0.037$), this primordial contribution is negligible so that the result can be interpreted as being solely due to nonlinearities at recombination.


\section{Conclusions}
\label{sec:conclusions}

Although the calculation of the CMB anisotropies at second order is remarkably complicated, the CMB bispectrum in the squeezed limit, with one of the modes outside the Hubble radius at recombination, is given by the simple formula \eqref{final_formula}. This expression takes into account all the effects at recombination and must be supplemented by the additional contributions to the bispectrum at low redshift. As we stressed, the formula already takes into account the primordial non-Gaussianity produced by any single-field model of inflation, at any order in slow-roll. From this point of view, eq.~\eqref{final_formula} can be seen as an extension of Maldacena's consistency relation \cite{Maldacena:2002vr,Creminelli:2004yq,Cheung:2007sv} in terms of observable quantities: a deviation of the temperature bispectrum from our eq.~\eqref{final_formula} (once low-redshift effects are properly subtracted) would rule out all single-field models. 

Of course, to fully exploit the forthcoming Planck data, one wants to consider also configurations of the bispectrum for which  \eqref{final_formula} does not apply. In this case one has to resort to a numerical Boltzmann code. Our simple formula can be used as a non-trivial check of the numerics, as all the second-order effects must exactly conspire to match our simple result in the squeezed limit. 
The current version of CMBquick is able to correctly reproduce our analytic results, though some residual disagreement indicates that some effort should be put to improve the numerical reliability of the code.

A further handle to constrain primordial non-Gaussianity will come from polarization measurements \cite{Babich:2004yc}. Our analysis can be extended to bispectra including polarization, along the lines indicated by \cite{Creminelli:2004pv}. We leave this for future work.

\vspace{0.5cm}
{\bf Acknowledgements:}
It is a pleasure to thank Guido D'Amico and Jorge Nore\~na for collaboration in the early stages of this project and Anthony Lewis, Alberto Nicolis, Uros Seljak and Matias Zaldarriaga for useful discussions. When completing this work, we have become aware of a similar work by N.~Bartolo, S.~Matarrese and A.~Riotto. Our results, when overlap is possible, agree with theirs. Note however that in their paper lensing is not considered. We thank them for useful correspondence.

\

\appendix


\section{CMB anisotropies at second order}
\label{app:CMB}

In this appendix we provide a second-order expression for the CMB temperature anisotropies in the perfect fluid approximation and assuming instantaneous recombination.
We are interested in computing the CMB temperature fluctuations
\be
\Theta_{\rm obs} (\hat n) \equiv \frac{T_{\rm obs}(\hat n) - \langle T_{\rm obs} \rangle}{\langle T_{\rm obs} \rangle}\;, \label{Theta}
\ee
where $T_{\rm obs}(\n)$ is the observed photon temperature in the angular
direction $\n$ ($\n^2 =1$) and $\langle T_{\rm obs} \rangle$ is its average over the
sky. We concentrate on the limit of instantaneous recombination so that the decoupling takes place at a given physical time $\eta_{\rm rec}$. 
For a black-body spectrum the observed temperature $T_{\rm obs}(\n)$ is
related to the one of emission $T_{\rm rec}$ by Liouville's theorem: as
phase space density is conserved in the propagation of photons, the phase space
density observed in a given direction $\n$ is the same as at emission but with
a rescaled temperature,
\be
T_{\rm obs}(\hat n) = \frac{E_{\rm obs}}{E_{\rm rec}} T_{\rm rec} (\eta_{\rm rec}, \vec x_{\rm rec})\;, \label{T_o}
\ee
where $E_{\rm rec}$ and $E_{\rm obs}$ are the energies at emission and
observation of a given photon. This equation is exact and
therefore holds at any order in perturbation theory.

We consider a flat FRW metric perturbed at second-order in Poisson gauge \cite{Bertschinger:1993xt},
\be
ds^2 = a^2 (\eta) \left[ - e^{2 \Phi} d \eta^2 + 2 \omega_i dx^i d\eta  + \left(e^{- 2 \Psi } \delta_{ij} + \gamma_{ij}  \right) dx^i dx^j \right] \;, \label{metric}
\ee
where $\omega_i$ is transverse, $\partial^i \omega_i=0$, and $\gamma_{ij}$ is transverse and traceless, $\partial^i \gamma_{ij}= 0 = \gamma^i_i$. We only consider scalar {\em primordial} perturbations. In this case the vector and tensor components $\omega_i$ and $\gamma_{ij}$, are only second-order quantities.

The energy of a photon as measured by an observer with four-velocity $u^\mu$, with normalization condition $u^\mu u_\mu = -1$, is given by
\be
E = - P_\mu u^\mu \;, \label{freq}
\ee
where $P^\mu \equiv dx^\mu /d \lambda$ is the four-momentum of the photon, $P^\mu P_\mu =0$. 
Thus, from eq.~\eqref{freq} we obtain
\be
\frac{E_{\rm obs}}{E_{\rm rec}} = \frac{a_{\rm rec} P_0(\eta_{\rm obs}) }{a_{\rm obs} P_0(\eta_{\rm rec}) } e^{\Phi_{\rm rec}} \left( \sqrt{1+v^2} + e^{\Phi+\Psi} \frac{P_i}{P_0} v^i \right)_{\rm rec}^{-1}\;, \label{omega_ratio}
\ee
where we have used $u^0 = \sqrt{ 1+ v^2} \; e^{-\Phi}/a$ from the normalization condition of the four-velocity and the definition of $v^i$ given above, $v^i = a e^{-\Psi} u^i$. We have also set to zero the velocity of the observer, which is simply responsible for a dipole.

The photon geodesic equation can be conveniently rewritten as \cite{Mukhanov:2005sc}
\begin{equation}
  \frac{d  P_\mu}{d \lambda} = \frac{1}{2}\partial_\mu g_{\alpha\beta}P^\alpha
  P^\beta \, . \label{geod}
\end{equation}
In order to compute $P_0$ we need to solve the
time component of eq.~\eqref{geod} up to second order. After few manipulations we can rewrite this, up to second order, as
\begin{equation}
\frac{1}{P_0}  \frac{d P_0}{d \eta} = \Phi' + \Psi' - \omega_i'\hat{n}^i -
  \frac{1}{2}\gamma_{ij}'\hat{n}^i\hat{n}^j \;,
\end{equation}
where we have used that $\n^i =   P^i/P^0=$ constant along the unperturbed geodesic.
Upon integration, this equation yields
\begin{equation}
\frac{P_0(\eta_{\rm obs})}{P_0(\eta_{\rm rec})}  =  \exp\left[ I(\eta_{\rm rec},\eta_{\rm obs}) \right]\;, \quad I(\eta_{\rm rec},\eta_{\rm obs}) \equiv \int_{\eta_{\rm rec}}^{\eta_{\rm obs}} d\eta\;\Big(\Phi' +
  \Psi' - \omega_i'\hat{n}^i -
  \frac{1}{2}\gamma_{ij}'\hat{n}^i\hat{n}^j\Big)\,.
\label{geodesic}
\end{equation}
Note that $P_0$ is conserved at zeroth order and we choose $P_0 = 1$.

As $v^i$ in eq.~\eqref{omega_ratio} is a first-order quantity, we need to find $P_i(\eta_{\rm rec})/P_0(\eta_{\rm rec})$ up to first order only. Thus, we solve the spatial component of eq.~\eqref{geod}. At first order this reads
\begin{equation}
\frac{1}{P_0}  \frac{d P_i}{d \eta} = \partial_i (\Phi + \Psi)\; ,
\end{equation}
which can be integrated 
using the zeroth-order value for $P_0$ giving
\be
P_i (\eta_{\rm rec}) =   P_i (\eta_{\rm obs})+  \int_{\eta_{\rm rec}}^{\eta_{\rm obs}} d\eta \;\partial_i (\Phi+\Psi) \;. \label{P_i}
\ee
Furthermore, eq.~\eqref{geodesic}  at first order reads
\be
P_0 (\eta_{\rm rec}) =  P_0 (\eta_{\rm obs}) -  \int_{\eta_{\rm rec}}^{\eta_{\rm obs}} d\eta \;(\Phi'+  \Psi') \;. \label{P_0_1st}
\ee
The observer photon direction is  $\n^i =  P_{\rm obs}^i/P_{\rm obs}^0 = - P_i(\eta_{\rm obs})/P_0(\eta_{\rm obs})$. Here we have set to zero the metric perturbations at the observer position: since metric perturbations at the observer do not depend on the direction of observation, they can be reabsorbed in the redefinition of the average temperature. 
The two equations above can be combined to give
\be
\frac{P_i (\eta_{\rm rec})}{P_0 (\eta_{\rm rec})} = -  \n^i (1-\Phi_{\rm rec} - \Psi_{\rm rec}) + \delta n^i (\eta_{\rm rec},\eta_{\rm obs}) \;, \quad  \delta n^i (\eta_{\rm rec},\eta_{\rm obs}) =  \int_{\eta_{\rm rec}}^{\eta_{\rm obs}} d\eta \;\nabla^\perp_i (\Phi+\Psi) \;, \label{geodesic_i}
\ee
where we have rewritten the spatial derivative in \eqref{P_i} as a spatial gradient orthogonal to the line of sight, $\nabla^\perp_i \equiv (\delta_{ij} - \hat n_i \hat n_j) \partial_j $, while the derivative along the line of sight gives a boundary term and an integral of $\Phi'+\Psi'$ which cancels with the last term of eq.~\eqref{P_0_1st}.

Using eqs.~\eqref{geodesic} and \eqref{geodesic_i}, eq.~\eqref{omega_ratio} becomes, up to second order,
\be
\frac{E_{\rm obs}}{E_{\rm rec}} =  \frac{a_{\rm rec}}{a_{\rm obs}} \exp\left[ \Phi_{\rm rec} + I(\eta_{\rm rec},\eta_{\rm obs})\right] \left( \sqrt{1+v_{\rm rec}^2} - \n \cdot \vec v_{\rm rec} + \vec{\delta n} (\eta_{\rm rec},\eta_{\rm obs}) \cdot \vec v_{\rm rec} \right)^{-1}\;.
\ee
Finally, plugging this expression into the expression for the temperature, eq.~\eqref{T_o}, using the definition of temperature perturbation, eq.~\eqref{Theta}, and expanding up to second order, we obtain
\be
\begin{split}
\Theta_{\rm obs}(\n) &= \Theta_{\rm rec} + \Phi_{\rm rec} + \n \cdot \vec v_{\rm rec} +  I(\eta_{\rm rec},\eta_{\rm obs})+ \Phi_{\rm rec} \Theta_{\rm rec}  + \frac12 \Phi_{\rm rec}^2  - \frac12 v_{\rm rec}^2  \\ &+ (\n \cdot \vec v_{\rm rec} + I(\eta_{\rm rec},\eta_{\rm obs})) (\Theta_{\rm rec} + \Phi_{\rm rec} + \n \cdot \vec v_{\rm rec})     - \vec{\delta n}(\eta_{\rm rec},\eta_{\rm obs}) \cdot \vec v_{\rm rec} + \frac12 I(\eta_{\rm rec},\eta_{\rm obs})^2 \;. \label{Theta_expression}
\end{split}
\ee
It is straightforward to verify that this expression agrees with the one give in \cite{Bartolo:2004ty}.

For the sake of clarity we choose to simplify this expression even further by assuming pure matter dominance and absence of anisotropic stresses. 
In matter dominance $I(\eta_{\rm rec},\eta_{\rm obs})$ is only second order since at first order $\Phi = \Psi =$ const. Moreover, this second-order contribution (which includes the Rees-Sciama effect, vector and tensor modes) does not affect the CMB bispectrum  in the squeezed limit  \cite{Boubekeur:2009uk}, so that we can neglect all terms involving $I$ in eq.~\eqref{Theta_expression}. Furthermore, as $\vec v$ vanishes in the large scale limit, we can neglect in eq.~\eqref{Theta_expression} all terms which are quadratic in $\vec v$. It is easy to realize that also the term $- \delta \vec n \cdot \vec v_{\rm rec}$ vanishes in the squeezed limit and can thus be neglected.\footnote{As $\vec  v_{\rm rec}$ vanishes at large scales, one has to take $\delta \vec n$ on the long mode. The long-wavelength modulation of $\delta \vec n$ affects the short-scale 2-point function as 
\be
\langle \vec v(\vec x_1) \Theta_{\rm obs} (\vec x_2)\rangle \cdot \delta \vec n(\vec x_1) \;.
\ee
If $\delta \vec n$ is constant, i.e.~the same at the two points $\vec x_1$ and $\vec x_2$, there is no effect as there is a cancellation with the contribution obtained exchanging the two points. Therefore, this effect comes only from the variation of the long-wavelength $\delta \vec n$ over the short distance $\vec x_1 - \vec x_2$ and this is suppressed in the squeezed limit. Notice, for comparison, that the same conclusions does not apply to lensing, that we will discuss below: in this case the variation of the lensing angle multiplies the {\em derivative} of the 2-point function of the short modes so that the result is not suppressed in the squeezed limit \cite{Boubekeur:2009uk}.
} \footnote{In general, the temperature at emission will not be isotropic, but will depend on the angle of emission $-\hat n + \delta \vec n$. Expanding $\Theta_{\rm rec} $ around the unperturbed photon emission angle introduces another second-order term, $- \delta \vec n \cdot  \nabla_{\hat n} \Theta_{\rm rec} (\hat n ) $ \cite{Pyne:1995bs}. By the same argument used for $- \delta \vec n \cdot \vec  v_{\rm rec}$, one can show that also this term vanishes in the squeezed limit. }
 This leaves us with
\be
\Theta_{\rm obs}(\n) = \left[ \Theta + \Phi + \n \cdot \vec v + \Phi \Theta +  \Phi^2/2 + (\Theta + \Phi) \n \cdot \vec v  \right] (\eta_{\rm rec},\vec x_{\rm rec})\;. \label{CMB2}
\ee

Note that the right-hand side of this equation depends on the physical time $\eta_{\rm rec}$ and position $\vec x_{\rm rec}$ of photon emission, which are perturbed quantities. Expanding around the unperturbed emission time and position gives rise to other second-order effects, well described in \cite{Pyne:1995bs}. We discuss these effects in more details in the main text.

\end{document}